\documentclass[sn-nature]{sn-jnl}

\usepackage{graphicx}%
\usepackage{multirow}%
\usepackage{amsmath,amssymb,amsfonts}%
\usepackage{amsthm}%
\usepackage{mathrsfs}%
\usepackage[title]{appendix}%
\usepackage{xcolor}%
\usepackage{textcomp}%
\usepackage{manyfoot}%
\usepackage{booktabs}%
\usepackage{algorithm}%
\usepackage{algorithmicx}%
\usepackage{algpseudocode}%
\usepackage{listings}%
\usepackage{siunitx}
\usepackage{float}
\usepackage{makecell}
\usepackage{lineno}

\theoremstyle{thmstyleone}%
\theoremstyle{thmstyletwo}%

\theoremstyle{thmstylethree}%
\begin{document}

\title[Article Title]{
OmniESI: A unified framework for enzyme-substrate interaction prediction with progressive conditional deep learning
}

\author[1,2]{\fnm{Zhiwei} \sur{Nie}}
\equalcont{These authors contributed equally to this work.}
\author[1]{\fnm{Hongyu} \sur{Zhang}}
\equalcont{These authors contributed equally to this work.}

\author[3]{\fnm{Hao} \sur{Jiang}}
\author[4]{\fnm{Yutian} \sur{Liu}}
\author[2]{\fnm{Xiansong} \sur{Huang}}
\author[2]{\fnm{Fan} \sur{Xu}}
\author[5]{\fnm{Jie} \sur{Fu}}
\author[2]{\fnm{Zhixiang} \sur{Ren}}

\author*[1,4,2]{\fnm{Yonghong} \sur{Tian}}\email{yhtian@pku.edu.cn}
\author*[3]{\fnm{Wen-Bin} \sur{Zhang}}\email{wenbin@pku.edu.cn}
\author*[1,2]{\fnm{Jie} \sur{Chen}}\email{jiechen2019@pku.edu.cn}

\affil[1]{School of Electronic and Computer Engineering, Peking University, Shenzhen, China}
\affil[2]{Pengcheng Laboratory, Shenzhen, China}
\affil[3]{Beijing National Laboratory for Molecular Sciences, Key Laboratory of Polymer Chemistry \& Physics of Ministry of Education, Center for Soft Matter Science and Engineering, College of Chemistry and Molecular Engineering, Peking University, China}
\affil[4]{School of Computer Science, Peking University, China}
\affil[5]{Shanghai AI Laboratory, Shanghai, China}

\abstract{
Understanding and modeling enzyme-substrate interactions is crucial for catalytic mechanism research, enzyme engineering, and metabolic engineering.
Although a large number of predictive methods have emerged, they do not incorporate prior knowledge of enzyme catalysis to rationally modulate general protein-molecule features that are misaligned with catalytic patterns.
To address this issue, we introduce a two-stage progressive framework, OmniESI, for enzyme-substrate interaction prediction through conditional deep learning.
By decomposing the modeling of enzyme-substrate interactions into a two-stage progressive process, OmniESI incorporates two conditional networks that respectively emphasize enzymatic reaction specificity and crucial catalysis-related interactions, facilitating a gradual feature modulation in the latent space from general protein-molecule domain to catalysis-aware domain.
On top of this unified architecture, OmniESI can adapt to a variety of downstream tasks, including enzyme kinetic parameter prediction, enzyme-substrate pairing prediction, enzyme mutational effect prediction, and enzymatic active site annotation. 
Under the multi-perspective performance evaluation of in-distribution and out-of-distribution settings, OmniESI consistently delivered superior performance than state-of-the-art specialized methods across seven benchmarks.
More importantly, the proposed conditional networks were shown to internalize the fundamental patterns of catalytic efficiency while significantly improving prediction performance, with only negligible parameter increases (0.16\%), as demonstrated by ablation studies on key components. 
Overall, OmniESI represents a unified predictive approach for enzyme-substrate interactions, providing an effective tool for catalytic mechanism cracking and enzyme engineering with strong generalization and broad applicability.
}

\maketitle

\section{Introduction}

Enzyme-substrate interactions (ESIs) are fundamental to biochemical reactions, forming the basis of metabolic pathways and signal transduction in organisms \cite{wolfenden1999conformational, ferrall2020reassessing}. 
Understanding these interactions is crucial for evaluating catalytic efficiency \cite{albery1976evolution} and determining the catalytic potential of enzyme-substrate pairs \cite{black2015high}.
Consequently, precise ESI modeling has become central to fields such as enzyme engineering \cite{campbell2016role,newton2018enzyme}, metabolic engineering \cite{sweetlove2018role}, and disease mechanism research \cite{ryan1989specificity}.
Traditionally, ESI analysis relies on wet-lab experimental measurements, which are time-consuming and costly, greatly limiting the feasibility of high-throughput analysis \cite{black2015highthrough1, longwell2017highthrough2}.
This bottleneck highlights the pressing need for efficient computational methods, spurring significant research efforts into various facets of ESI modeling.

Current computational methods for ESI prediction can be broadly classified into two categories based on distinct training paradigms: methods trained from scratch and those leveraging pretrained models. 
Methods trained from scratch \cite{kroll2021kroll_km, li2022dlkcat, du2023gelkcat, qiu2024dltkcat} directly utilized the ESI database to train all model components with randomly initialized parameters, including the specific encoders for both enzyme and substrate, as well as the fusion module.
In contrast, methods leveraging pretrained models employed protein and molecule foundation models to extract embeddings for enzymes and their substrates, which were then applied to diverse prediction tasks, such as predicting enzyme kinetic parameters \cite{kroll2023turnup, yu2023unikp, he2024graphkm, kroll2024prosmith, wang2024mpek,wang2025robust, boorla2025catpred}, determining whether a small molecule is a substrate of an enzyme \cite{kroll2023esp,kroll2024prosmith}, predicting the fitness effects upon mutations \cite{yu2023unikp}, and annotating enzymatic active site \cite{wang2024multi}.
Benefiting from the powerful capabilities of pretrained protein and molecule models in various tasks \cite{suzek2015uniref, meier2021esm1v, lin2023esm, elnaggar2021prottrans,chithrananda2020chemberta, wang2022molclr, xiamole-bert}, ESI predictive methods leveraging pretrained models have been demonstrated to outperform those trained from scratch, which makes them the current mainstream computational approaches for ESI research.

However, existing methods directly employed initial-extracted enzyme and substrate embeddings \cite{kroll2023turnup, yu2023unikp, he2024graphkm, kroll2024prosmith, wang2024mpek, wang2025robust, boorla2025catpred, kroll2023esp} or simply implemented feature interaction based on attention mechanism or adaptive gate network \cite{li2022dlkcat, wang2024multi, du2023gelkcat,qiu2024dltkcat}, without introducing prior knowledge of enzyme catalysis to reasonably modulate the features of enzyme and its substrate.
Specifically, as illustrated in Fig.\ref{fig1}a, the pretrained protein and molecule encoders or those trained from scratch map enzymes and substrates into the latent space of general protein-molecule features due to the lack of explicit guidance from catalysis-related priors, which are misaligned with enzyme-specific, substrate-specific, and catalysis-aware features.
This mismatch can lead to the neglect of crucial interaction-related information, such as enzymatic active sites and functional groups of substrate, during the modeling of enzyme-substrate interactions, while also introduces catalysis-irrelevant noise for neural networks.
Therefore, a reasonable feature modulation strategy guided by the enzyme-substrate interaction patterns is urgently needed.

To solve this problem, we introduce a two-stage progressive framework, OmniESI, for enzyme-substrate interaction modeling through conditional deep learning.
As shown in Fig.\ref{fig1}b, the process of ESI modeling is conceptualized as a two-stage progressive feature modulation using two conditional networks, starting with a feature transition to enzyme-substrate domain, followed by a final shift to catalysis-aware domain.
First, due to the inherent specificity of enzymatic reactions \cite{gottschalk1947enz_special}, an enzyme and its substrate naturally serve as reciprocal conditional information for one another.
This reciprocal conditioning facilitates the feature modulation from general protein-molecule domain to specialized enzyme-substrate domain, actively emphasizing the enzymatic reaction specificity.
Second, the enzyme-substrate interaction itself can also act as conditional information, enabling the further feature transition from the enzyme-substrate domain to catalysis-aware domain.
This shift actively underscores the significance of specific enzyme residues and substrate atoms that are responsible for forming crucial molecular interactions, such as hydrogen bonds \cite{wang2008hydro} and $\pi-\pi$ stacking interactions \cite{gong2021pai}, which are essential for catalytic process.
After the above two-stage feature modulation, the final catalysis-aware features that encapsulate fine-grained ESI patterns can be applied to different types of catalysis-related downstream tasks, namely enzyme kinetic parameter prediction, enzyme-substrate pairing prediction, enzyme mutational effect prediction, and enzymatic active site annotation (Fig.\ref{fig1}c).

Through two conditional networks with different tendencies of feature modulation, OmniESI was demonstrated to be highly applicable to different types of downstream tasks in this work.
We showcased the predictive power of OmniESI under both in-distribution (ID) and out-of-distribution (OOD, decreasing enzyme sequence identity to training sequences) evaluation protocols for comprehensive performance evaluation.
OmniESI was first shown to accurately predict enzyme kinetic parameters, namely the turnover number ($k_{cat}$), the Michaelis constant ($K_m$), and the inhibition constant ($K_i$).
OmniESI was then demonstrated to predict enzyme-substrate pairs with high performance, a prediction task to mine potential substrates.
Next, OmniESI was shown to accurately annotate enzymatic active sites and internalize catalytic reaction mechanisms via attention weights.
Finally, OmniESI's predictive power was evaluated on enzyme mutational effect prediction tasks, achieving significant performance improvements on both single-point mutation and epistasis-related double-point mutation scenarios.
Considering all downstream tasks, OmniESI surpassed the state-of-the-art specialized methods on most of evaluation metrics, and was the best or second best on all of them.
Ablation experiments from both quantitative and qualitative perspectives confirmed that the designed conditional networks only introduced negligible number of parameters but was significantly beneficial to performance improvements of OmniESI.

\section{Results}

\subsection{OmniESI architecture}

\begin{figure}[h!]%
\centering
\includegraphics[width=\linewidth]{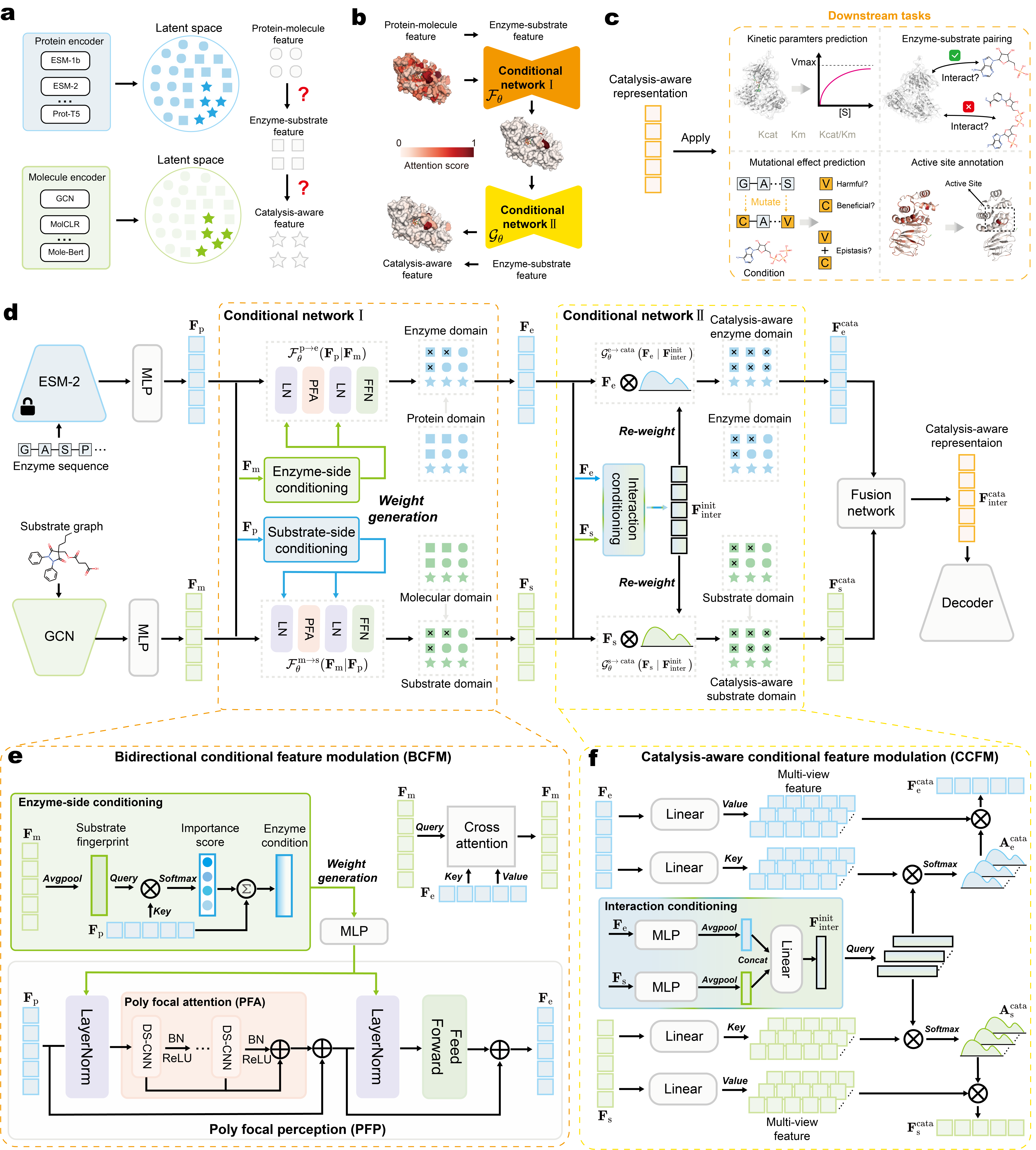}
\caption{
The methodology and architecture of OmniESI.
\textbf{a}, The illustration of OmniESI's motivation, where various encoders map enzymes and substrates into the latent space of general protein-molecule features, leading to misalignment with enzyme-specific, substrate-specific, and catalysis-aware features.
\textbf{b}, The methodology of OmniESI, where a two-step progressive feature modulation is performed through two conditional networks to modulate the initial-extracted features into catalysis-aware domain.
\textbf{c}, Downstream tasks of OmniESI, namely enzyme kinetic parameter prediction, enzyme-substrate pairing prediction, enzyme mutational effect prediction, and enzymatic active site annotation.
\textbf{def}, The architecture of OmniESI, which consists of four modules: enzyme-substrate encoding, bidirectional conditional feature modulation (BCFM, e), catalysis-aware conditional feature modulation (CCFM, f), and task-specific decoding.
}
\label{fig1}
\end{figure}

As illustrated in Fig.\ref{fig1}d, OmniESI harnesses a progressive conditional architecture, which consists of four modules: enzyme-substrate encoding, bidirectional conditional feature modulation (BCFM, i.e. conditional network I), catalysis-aware conditional feature modulation (CCFM, i.e. conditional network II), and task-specific decoding.

First, for the enzyme-substrate encoding module, the amino-acid sequence of enzyme and the 2D molecular graph of substrate are extracted for embeddings by their corresponding encoders.
We adopt the widely recognized ESM-2 (650M) \cite{lin2023esm} with frozen parameters for enzyme sequence and the graph convolutional network (GCN) \cite{kipf2016gcn} trained from scratch for substrate 2D graph.

Second, the extracted embeddings of enzyme and substrate are fed into the conditional network I, named bidirectional conditional feature modulation (BCFM) here, in parallel.
This conditional network is operated in a two-way manner: the enzyme serves as the condition for the substrate-side network, while the substrate acts as the condition for the enzyme-side network.
Such reciprocal conditioning allows the enzyme-side and substrate-side networks to actively capture catalytic specificity information, thereby modulating the features from the general protein-molecule domain to the enzyme-substrate domain.
Specifically, as shown in Fig.\ref{fig1}e, the conditional network I includes two key blocks: the poly focal perception (PFP) block and the two-sided conditioning (TC) block.
PFP block extracts fine-grained contextual representation of enzymes and substrates across diverse receptive fields through multiple large kernel depthwise separable convolutions (DSCNN) \cite{chollet2017depthwise,yu2024inceptionnext, cai2024pki}.
Meanwhile, TC block is designed to generate enzyme-side and substrate-side conditional embeddings, which directs PFP block to internalize the information from its catalytic partner through weight generation.
In addition, multi-head cross-attention \cite{chen2021crossvit} is adopted to update the features of enzyme and substrate serving as the conditions after the forward process of PFP block on each side.
In summary, following a bidirectional symmetric manner, enzyme or substrate is used to support the feature modulation process of its catalytic partner.

Third, the modulated features of enzyme and substrate enter the conditional network II, named catalysis-aware conditional feature modulation (CCFM) here, in parallel.
This conditional network derives the initial joint representation of an enzyme and its substrate as a condition to facilitate the feature modulation from the enzyme-substrate domain to catalysis-aware domain.
Specifically, as shown in Fig.\ref{fig1}f, the representations of an enzyme and its substrate modulated by the conditional network I are concatenated into an initial aggregated representation, which represents a rough enzyme-substrate interaction pattern.
Subsequently, taking the above aggregated representation as condition embedding (i.e. query vector), the features of enzymes and their substrates re-weighted to obtain fine-grained catalysis-aware features.

Fourth, the catalysis-aware features of enzymes and their substrates are integrated into a final representation for task-specific decoding.
In this work, four representative catalysis-related downstream tasks are covered by OmniESI, namely enzyme kinetic parameter prediction, enzyme-substrate pairing prediction, enzyme mutational effect prediction, and enzymatic active site annotation, as illustrated in Fig.\ref{fig1}c.

\subsection{Prediction performance for enzyme kinetic parameters and enzyme-substrate pairs}

\begin{figure}[h!]%
\centering
\includegraphics[width=\linewidth]{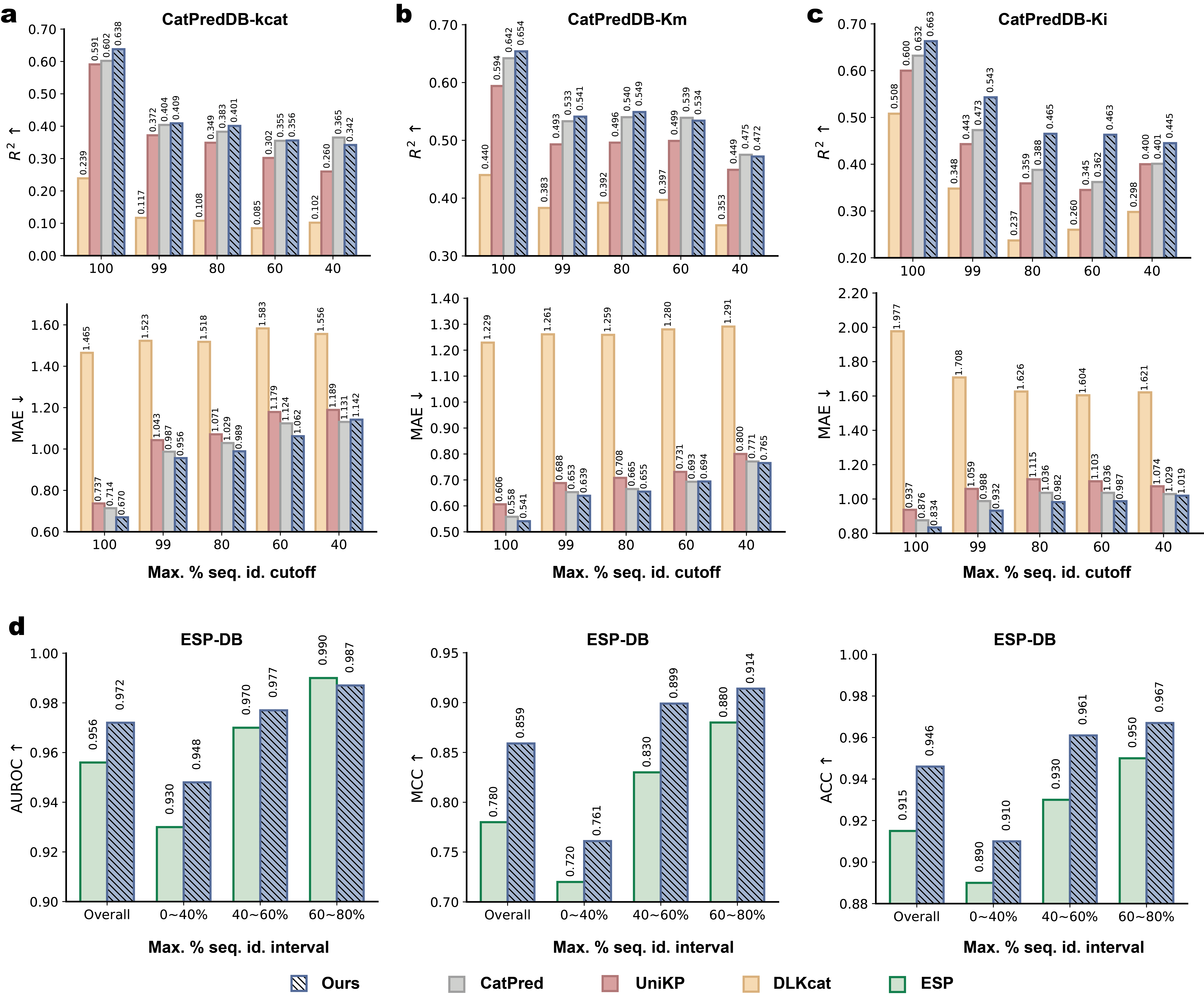}
\caption{
Performance evaluation for enzyme kinetic parameters and enzyme-substrate pairs.
An upward arrow means higher values are better, while a downward arrow means lower values are better.
\textbf{abc}, Performance evaluation under in-distribution and out-of-distribution setting for enzyme kinetic parameters, namely $k_{cat}$ (a), $K_m$ (b), and $K_i$ (c).
The coefficient of determination ($R^2$) and mean absolute error (MAE) are plotted.
"Max. \% seq. id. cutoff" denotes the maximum percentage sequence identity of test subsets to training sequences, and that of 100 refers to in-distribution evaluation.
\textbf{d}, Performance evaluation for enzyme-substrate pairs, where area under the receiver operating characteristic curve (AUROC), matthews correlation coefficient (MCC), and accuracy (ACC) are plotted.
"Max. \% seq. id. interval" denotes the interval of maximum percentage sequence identity of test subsets to training sequences, and "Overall" refers to the test set with enzyme sequence identity interval of 0-80\% to training sequences.
}
\label{fig2}
\end{figure}

The catalytic efficiency of enzymes for specific substrates is crucial for enzyme engineering \cite{sharma2021enzyme} and metabolic engineering \cite{woolston2013metabolic}, so it is very beneficial to develop predictive methods for enzyme kinetic parameters that are not dependent on experimental measurements.
Moreover, there are large gaps in the mapping between most enzymes and the substrates catalyzed by them, which hinders the efficient mining of potential substrates.
Therefore, it is also of great value to develop more accurate enzyme-substrate pairing (i.e., determining whether a small molecule is a substrate of an enzyme) predictive methods.
Accordingly, in this subsection, the prediction performance of OmniESI was comprehensively evaluated for the above two downstream tasks, namely enzyme kinetic parameter prediction and enzyme-substrate pairing prediction.

For enzyme kinetic parameter prediction tasks, the turnover number ($k_{cat}$), the Michaelis constant ($K_m$), and inhibition constant ($K_i$) that describes the affinity of the inhibitor binding to the enzyme are adopted as the prediction targets of regression tasks.
The compared baseline models are DLKcat \cite{li2022dlkcat}, UniKP \cite{yu2023unikp}, and CatPred \cite{boorla2025catpred}.
The curated CatPred-DB datasets \cite{boorla2025catpred} were adopted as the benchmark datasets for enzyme kinetic parameter prediction tasks.
In this work, we used the raw data splits of CatPred-DB literature \cite{boorla2025catpred} for fair comparison (Methods \ref{method_kinetic_data}).
Specifically, in CatPred-DB, the held-out test sets for $k_{cat}$, $K_m$, and $K_i$, allowed a maximum percentage sequence identity (Max. \% seq. id. cutoff) to training sequences of 100\%, and were therefore used here for in-distribution (ID) evaluation. 
Using different sequence identity cutoff values (99\%, 80\%, 60\%, and 40\%), the above ID test sets were further filtered into multiple subsets for out-of-distribution (OOD) performance evaluation.
For $k_{cat}$ prediction in Fig.\ref{fig2}a and Supplementary Table \ref{tables:kcat}, OmniESI comprehensively outperformed state-of-the-art methods under ID performance evaluation, improving the coefficient of determination ($R^2$) by 6.0\% and mean absolute error (MAE) by 6.2\% relative to CatPred model.
On four OOD test subsets with different sequence identities, OmniESI lagged behind CatPred model only when set Max. \% seq. id. cutoff was 40\%, and outperformed other methods on the remaining three subsets.
OmniESI's performance for $K_m$ prediction was similar to that for $k_{cat}$ prediction, among which OmniESI showed a more superior performance under ID performance evaluation, and outperformed other methods in two of the four OOD test subsets (Fig.\ref{fig2}b and Supplementary Table \ref{tables:km}).
It is worth noting that OmniESI showed significant performance advantages for $K_i$ prediction, outperforming other methods comprehensively in both ID and OOD performance evaluations (Fig.\ref{fig2}c and Supplementary Table \ref{tables:ki}).
Under ID performance evaluation, OmniESI achieved relative improvements of 4.9\% in $R^2$ and 4.8\% in MAE.
Under OOD performance evaluation, $R^2$ and MAE metrics of OmniESi were greater than those of its competitors, by an average of 18.4\% and 5.4\% respectively.
In summary, based on the results of above three prediction targets, OmniESI consistently outperformed state-of-the-art methods in ID performance evaluation, and demonstrated better performance in most sequence identity settings when conducting OOD performance evaluation.

For enzyme-substrate pairing prediction tasks, the state-of-the-art method ESP \cite{kroll2023esp} was adopted for performance comparison with OmniESI and the curated ESP database (ESP-DB) \cite{kroll2023esp} was adopted as the benchmark dataset for enzyme-substrate pairing prediction task.
In this work, we used the raw data splits of ESP-DB literature \cite{kroll2023esp} for fair comparison (Methods \ref{method_pair_data}).
Specifically, in ESP-DB, the test set was designed for OOD evaluation, where the sequence identity of enzyme sequences in it didn't exceed 80\% to training sequences.
The above test set was further divided into three test subsets with different intervals of maximum percentage sequence identity (Max. \% seq. id. interval), namely 0-40\%, 40-60\%, and 60-80\%, for more comprehensive OOD evaluation.
As shown in Fig.\ref{fig2}d and Supplementary Table \ref{tables:pair_overall}, on the "overall" test set with maximum 80\% sequence identity to training sequences, OmniESI surpassed ESP across the board, improving area under the receiver operating characteristic curve (AUROC), matthews correlation coefficient (MCC), and accuracy (ACC) by 0.016, 0.079, and 3.1\%, respectively.
On the test subsets with different enzyme sequence identity intervals, OmniESI performed favorably against ESP model, especially for the enzyme sequence identity intervals 0-40\% and 40-60\% (Supplementary Table \ref{tables:pair_subset}).
For example, for the OOD performance evaluation with sequence identity between 40-60\%, the MCC and ACC metrics were improved by 0.069 and 3.1\%, respectively.
Overall, OmniESI is more competitive against existing prediction methods for enzyme-substrate pairs.

\subsection{Prediction performance for enzymatic active sites}

\begin{figure}[h!]%
\centering
\includegraphics[width=\linewidth]{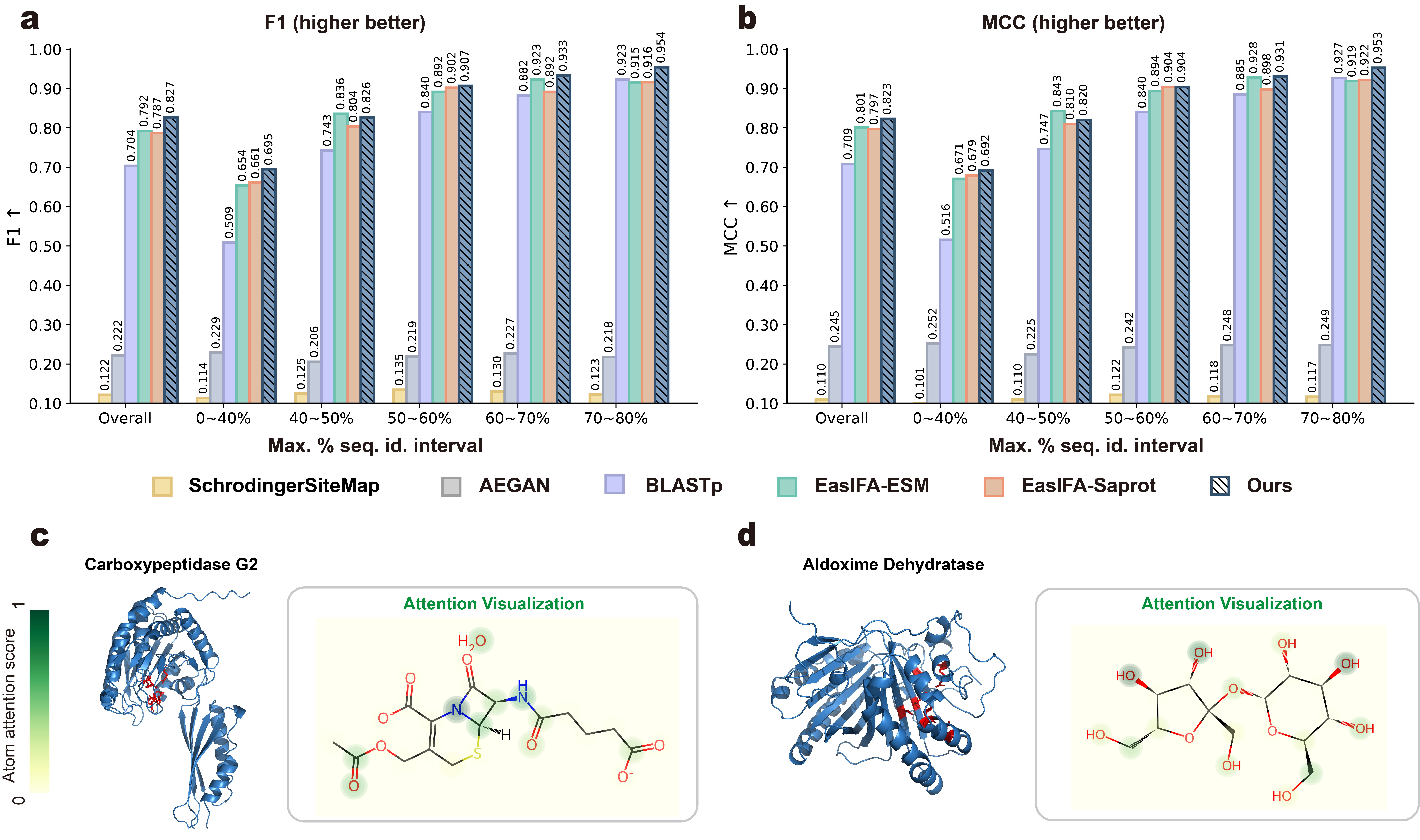}
\caption{
Performance evaluation for enzymatic active sites.
\textbf{ab}, The F1-score (a) and MCC (b) metrics are plotted for performance comparison.
An upward arrow means higher values are better.
"Max. \% seq. id. interval" denotes the interval of maximum percentage sequence identity of test subsets to training sequences, and "Overall" refers to the test set with enzyme sequence identity interval of 0-80\% to training sequences.
\textbf{cd}, Attention weight visualization for enzyme-substrate interaction on substrate atoms of two cases.
}
\label{fig3}
\end{figure}

The active site of an enzyme is the region that directly binds to the substrate specifically and catalyzes the chemical reaction against the substrate \cite{tsou1993conformational}, thereby determining the enzymatic activity \cite{toscano2007minimalist,osuna2015molecular}.
Considering the importance of enzymatic active site annotation for enzyme engineering, the prediction performance of OmniESI was evaluated for this downstream task.
The SwissProt E-RXN ASA dataset \cite{wang2024multi} was adopted as the benchmark dataset for enzymatic active site annotation task.
In this work, we used the raw data splits of SwissProt E-RXN ASA literature \cite{wang2024multi} for fair comparison (Methods \ref{method_site_data}).
Specifically, in SwissProt E-RXN ASA dataset, the test set was suitable for OOD evaluation, where the enzymes with more than 80\% sequence identity to training sequences were removed.
For more comprehensive OOD evaluation, the above test set was further filtered into five test subsets with different intervals of maximum percentage sequence identity (Max. \% seq. id. interval): 0-40\%, 40-50\%, 50-60\%, 60-70\%, and 70-80\%.

As shown in Fig.\ref{fig3}ab, a binary classification task of distinguishing whether a residue is an active site was set.
The compared baseline models are Schrodinger-SiteMap \cite{halgren2009identifying}, AEGAN \cite{shen2023highly}, BLASTp \cite{altschul1990basic}, EasIFA-ESM \cite{wang2024multi}, and EasIFA-SaProt \cite{wang2024multi}.
From the performance comparison results, we can find that OmniESI was more competitive than existing predictive approaches.
First, OmniESI surpassed state-of-the-art EasIFA-ESM model in both F1-score and MCC on the "overall" test set with enzyme sequence identity interval of 0-80\% to training sequences, achieving relative improvements of 4.4\% and 2.7\%, respectively (Supplementary Table \ref{tables:site_overall}).
Second, on the other five test subsets with different enzyme sequence identity intervals, OmniESI outperformed other baseline methods except for the sequence identity interval of 40-50\% (Supplementary Table \ref{tables:site_f1} and Table \ref{tables:site_mcc}).
Taking the F1-score metric as an example, OmniESI lagged slightly behind EasIFA-ESM on the test subset with a sequence identity interval of 40-50\%, but surpassed other methods on the remaining four test subsets, especially achieving a 5.1\% improvement relative to other methods for sequence identity interval of 0-40\%.

Further, to observe the interpretability of OmniESI's attention weights for catalytic interactions, we performed zero-shot inference of OmniESI (i.e., the model weights trained with SwissProt E-RXN ASA training set were directly used for testing) on high-quality annotated cases from the test set of MCSA E-RXN CSA dataset \cite{wang2024multi} (Methods \ref{method_site_data}).
The aggregated representation of an enzyme and its substrate in CCFM module was employed as the query vector to calculate the attention weights on substrate atoms, which was to highlight the important substrate atoms that contributed to the catalytic interactions.
Two representative enzyme-substrate catalytic reactions were used as case studies, namely amide hydrolysis in Fig.\ref{fig3}c and glycosylation in Fig.\ref{fig3}d.
In the case of amide hydrolysis, the C–N bond in the amide bond as the reaction center was highlighted, and water molecules were also paid attention to.
In the case of glycosylation, the hydroxyl groups involved in the formation of the glycosidic bond were highlighted.
The above visualization results consistent with the enzyme catalytic mechanism demonstrated that OmniESI had internalized the enzyme-substrate interaction to a certain extent.

\subsection{Prediction performance for enzyme mutational effects}

\begin{figure}[h!]%
\centering
\includegraphics[width=\linewidth]{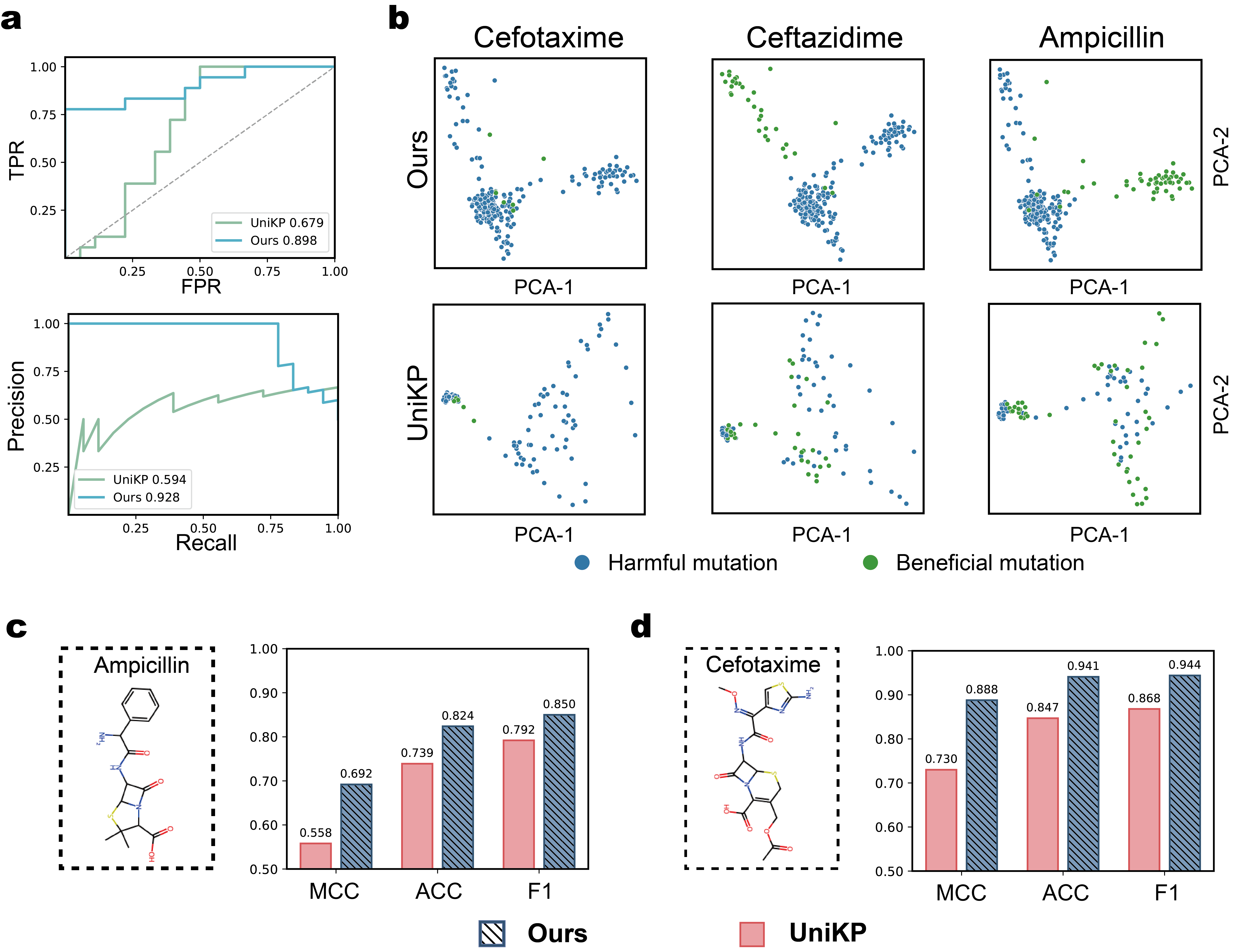}
\caption{
Performance evaluation for enzyme mutational effects.
\textbf{ab}, Quantitative and qualitative performance evaluation for enzyme single-point mutations.
The Receiver Operating Characteristic Curve and the Precision-Recall Curve are plotted for quantitative evaluation (a), where a binary classification task to distinguish harmful (decreased fitness) and beneficial mutations (increased fitness) was set.
The joint representations of three types of substrates (cefotaxime, ceftazidime, and ampicillin) and corresponding single-point enzyme mutants were visualized by dimensionality reduction via principal component analysis (PCA) for qualitative evaluation (b).
\textbf{cd}, Performance evaluation for enzyme double-point mutations against ampicillin (c) and cefotaxime (d), where a binary classification task to distinguish positive and negative epistasis was set.
}
\label{fig4}
\end{figure}

Understanding the relationship between amino-acid sequence and enzyme function helps to accelerate the directed evolution of enzyme for specific functions \cite{holliday2009understanding,song2023rational}, and the fitness effect upon mutations naturally becomes the middleman that bridges the sequence and function \cite{bartlett2002analysis,ribeiro2020global}.
For the evolutionary fitness landscape of enzymes, different types of mutations, namely beneficial mutations, harmful mutations, and neutral mutations, lead to different fitness effects, thereby affecting enzyme function, which is reflected in the changes in kinetic parameters such as $k_{cat}$ and $K_{m}$.
Furthermore, when multiple mutations occur, the extremely intricate interaction network between residues leads to epistasis \cite{starr2016epistasis}.
Epistasis means that the fitness effect caused by multiple mutations is not the additive fitness effect of the individual mutation involved \cite{miton2016mutational}.
Taking the case of double-point mutations as an example, when the fitness effect caused by double-point mutations is higher than the additive result, it is considered to be a positive epistasis dominated by synergistic interactions, otherwise it is a negative epistasis dominated by antagonistic interactions.
Accordingly, in this subsection, we showcased the predictive power of OmniESI for enzyme mutational effects in two scenarios, namely single-point mutations and double-point mutations involving epistasis.

$\beta$-lactamases pose a serious threat to the efficacy of $\beta$-lactam drugs \cite{drawz2010three}, the most widely prescribed class of antibiotics, and are one of the main culprits for bacterial resistance \cite{therrien2000molecular}.
Therefore, the CTX-M $\beta$-lactamase \cite{canton2006ctx,ambler1980structure,ambler1991standard}, a representative extended-spectrum $\beta$-lactamase, was adopted here for performance evaluation.
Specifically, a single-point mutational effect dataset and two double-point mutational effect datasets were curated from the deep mutational scanning (DMS) experiments of CTX-M $\beta$-lactamase of a previous work \cite{judge2023mapping} (Methods \ref{method_mutation_data}).
The DMS data of single-point mutations came from CTX-M $\beta$-lactamase against three types of substrates (i.e. $\beta$-lactam antibiotics, namely ampicillin, cefotaxime, and ceftazidime), while that of double-point mutations involved two types of substrates, namely ampicillin and cefotaxime.
It is worth noting that since the DMS data involved a large number of enzyme mutants that lack 3D structures, we compared the state-of-the-art sequence-based method UniKP with OmniESI in this subsection.

First, for enzyme single-point mutations, we set up a binary classification task to determine whether a single-point mutation is beneficial (increased fitness) or harmful (decreased fitness).
As shown in Fig.\ref{fig4}a and Supplementary Table \ref{tables:mutation_single}, OmniESI significantly outperformed UniKP in both AUPRC and AUROC, achieving 0.334 and 0.219 improvements, respectively.
The accuracy of 50\% and MCC of 0 indicated that the prediction of UniKP was close to random guessing, while OmniESI reached an accuracy of 86.1\% and a MCC of 0.752.
In addition to the above quantitative evaluation, we further conducted a qualitative evaluation, where the joint representations of three types of substrates (cefotaxime, ceftazidime, and ampicillin) and corresponding single-point enzyme mutants were visualized by dimensionality reduction via principal component analysis (PCA) in Fig.\ref{fig4}b.
We can find that there was a significant overlap between beneficial and harmful mutations in the dimensionality reduction results of UniKP, while OmniESI distinguished the two types of mutations with relatively clear boundaries.
Second, for enzyme double-point mutations, we also set up a binary classification task to determine whether the epistasis of double-point mutations is positive or negative.
For the ampicillin substrate, OmniESI comprehensively surpassed its competitor UniKP, improving MCC, ACC, and F1 score by 0.134, 8.5\%, and 0.058, respectively (Fig.\ref{fig4}c and Supplementary Table \ref{tables:mutation_amp}).
For the cefotaxime substrate, OmniESI's significant advantage still existed, achieving a MCC of 0.888, an ACC of 94.1\%, and a F1-score of 0.944 (Fig.\ref{fig4}d and Supplementary Table \ref{tables:mutation_cefo}).

\subsection{Ablation study on key components}

\begin{figure}[h!]%
\centering
\includegraphics[width=\linewidth]{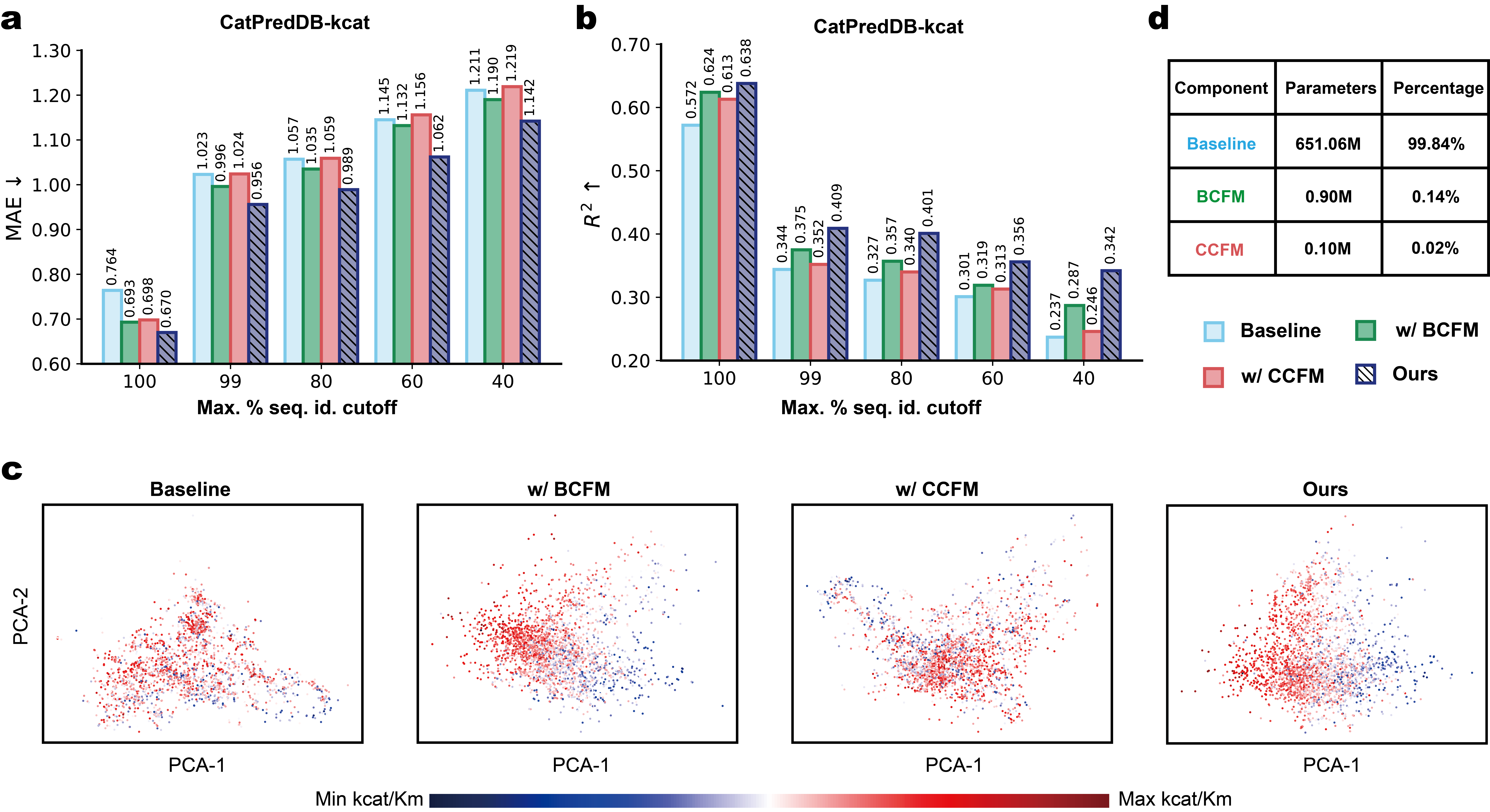}
\caption{
Ablation study on two key components, namely BCFM and CCFM module.
\textbf{ab}, Quantitative ablation experiments on BCFM and CCFM module, where MAE (a) and $R^2$ (b) are plotted for performance comparison under ID and OOD setting.
An upward arrow means higher values are better, while a downward arrow means lower values are better.
"Max. \% seq. id. cutoff" denotes the maximum percentage sequence identity of test subsets to training sequences, and that of 100 refers to ID performance evaluation.
\textbf{c}, Qualitative ablation experiments on BCFM and CCFM module, where the final aggregated representation of an enzyme and its substrate was adopted for dimensionality reduction visualization.
The enzyme-substrate pairs were divided into two categories with the median of $k_{cat}/K_m$ as the dividing line.
\textbf{d}, Parameter analysis of key components, where the parameter amounts of the individual components and their percentages are noted.
}
\label{fig5}
\end{figure}

There are two key components in OmniESI, bidirectional conditional feature modulation (BCFM) and catalysis-aware conditional feature modulation (CCFM), whose impacts on the predictive ability require a comprehensive ablation study from both quantitative and qualitative perspectives.
In this subsection, we created three model variants of OmniESI, namely Baseline (without either of the two modules), w/ BCFM (with BCFM module), and w/ CCFM (with CCFM module), to explore the contribution of each component to the prediction performance.

For the quantitative ablation experiments, the turnover number $k_{cat}$ prediction task was adopted for performance comparison using the $k_{cat}$ subset of CatPred-DB \cite{boorla2025catpred}.
As shown in Fig.\ref{fig5}ab and Supplementary Table \ref{tables:ablation}, under the ID performance evaluation (Max. \% seq. id. cutoff of 100\%), the introduction of any module improved the prediction performance of OmniESI, among which BCFM brought greater performance improvement compared with CCFM.
When both BCFM and CCFM existed, the prediction performance was further improved.
On four OOD test subsets with different sequence identity levels (99\%, 80\%, 60\%, 40\%), the ablation results were similar to those under the ID evaluation.
That is, any module was positive for performance improvement, and the joint use of the two modules further improved the prediction performance of OmniESI.

For the qualitative ablation experiments, the above models trained on the $k_{cat}$ dataset were used to distinguish different catalytic efficiency categories, where the median of $k_{cat}/K_m$ was adopted as the dividing line.
That is, if an enzyme-substrate pair had a $k_{cat}/K_m$ label greater than the median, it was considered to belong to the category of high-speed catalytic efficiency, otherwise it was the category of low-speed catalytic efficiency.
As shown in Fig.\ref{fig5}c, the final aggregated representations of enzymes and their substrates from three model variants and the complete OmniESI were visualized through dimensionality reduction via PCA.
We can find that the introduction of BCFM module significantly improved the discrimination between the two types of enzyme-substrate pairs, while the introduction of CCFM had mild effect.
In addition, the simultaneous introduction of the two modules further improved the distinction between the two types of enzyme-substrate pairs, resulting in a clear boundary between them.
The above qualitative ablation results were consistent with the quantitative ablation results, which was a double verification of the contribution of each key component.

Importantly, the key-component parameter analysis in Fig.\ref{fig5}d and Supplementary Table \ref{tables:para_analysis} demonstrated that the proposed two modules achieved significant performance improvements, but only introduced negligible parameter increases.
In the entire OmniESI framework, the BCFM module and the CCFM module only accounted for 0.14\% and 0.02\% of the parameters, but significantly improved the model's predictive ability both qualitatively and quantitatively.
This effectiveness of the proposed modules suggests that our two-stage progressive feature modulation strategy introduces a positive inductive bias to enhance the expressiveness of neural networks for enzyme-substrate interaction modeling tasks.

\section{Discussion}

In this work, the modeling of enzyme-substrate interactions is decomposed into a two-stage progressive process via two conditional networks.
By emphasizing the enzymatic reaction specificity and crucial catalytic interactions, general protein-molecule features are gradually modulated to be catalytically aware.
Therefore, OmniESI can adapt to different types of catalysis-related prediction targets with this unified framework.
Four important catalysis-related downstream tasks are adopted to evaluate the predictive ability and applicability of OmniESI, namely enzyme kinetic parameter prediction, enzyme-substrate pairing prediction, enzyme mutational effect prediction, and enzymatic active site annotation.
First, for enzyme kinetic parameter prediction tasks, the performance evaluation was conducted under both ID and OOD setting, where OmniESI consistently outperformed existing predictive methods under ID evaluation setting, and achieved better performance in most sequence identity settings under OOD performance evaluation.
Second, for enzyme-substrate pairing prediction tasks, OmniESI performed favorably against existing prediction methods on the overall OOD test set and three test subsets with different enzyme sequence identity intervals, especially for the identity intervals 0-40\% and 40-60\% that brought greater OOD challenges.
Third, for enzymatic active site annotation tasks, OmniESI outperformed other methods on the overall OOD test set and demonstrated superior performance on the five test subsets with different enzyme sequence identity intervals except for the sequence identity interval of 40-50\%.
In two case studies, OmniESI's attention weights showed alignment with the enzyme catalytic mechanism, indicating its internalization of enzyme-substrate interactions.
Fourth, for enzyme mutational effect prediction tasks, the predictive power of OmniESI was showcased for single-point mutations and double-point mutations involving epistasis, respectively.
Through quantitative and qualitative performance evaluation, OmniESI demonstrated comprehensive and significant performance improvements for both single-point and double-point mutational effect prediction scenarios compared with state-of-the-art methods.
In summary, OmniESI consistently demonstrated superior performance across different catalysis-related downstream tasks, achieving the best or second best on all evaluation metrics (outperforming existing predictive methods on most of them) under both in-distribution and out-of-distribution performance evaluation.
Furthermore, quantitative and qualitative ablation experiments demonstrated the positive contributions of two key components (BCFM module and CCFM module) of OmniESI, and component parameter analysis indicated that these key components introduced beneficial inductive biases to facilitate the modeling of enzyme-substrate interactions with negligible parameter increases.
Overall, OmniESI is a unified predictive method for enzyme-substrate interactions, showing strong generalizability and broad applicability in our multi-perspective evaluation.
OmniESI is applicable to a wide range of catalysis-related downstream tasks, and thus is expected to promote disease and catalytic mechanism research, enzyme engineering, metabolic engineering, etc.

Although OmniESI shows highly competitive out-of-distribution generalization capabilities compared to existing methods, its prediction performance gradually decreases as the test sequences become increasingly dissimilar to the training sequences.
On the one hand, as the number of enzyme-substrate interaction data publicly available in the community increases, incorporating these data into the training set may further improve OmniESI's predictive performance.
On the other hand, integrating specialized out-of-distribution generalization techniques into OmniESI is a more feasible goal in the short term.
Modifying OmniESI from the perspective of architecture (e.g., domain-invariant representation learning) or training strategy (e.g., meta-learning) is a potential direction that is expected to significantly improve its predictive ability for "unseen" enzyme-substrate interactions.

\section{Methods}
\subsection{Datasets}

\subsubsection{Enzyme kinetic parameter dataset}\label{method_kinetic_data}

The CatPred-DB datasets from a previous work \cite{boorla2025catpred} were adopted as the benchmark datasets for enzyme kinetic parameter prediction tasks.
These datasets contain the \textit{in~vitro} experimental measurements of $k_{cat}$, $K_m$, and $K_i$ that were curated from BRENDA database \cite{chang2021brenda} and SABIO-RK database \cite{wittig2018sabio}.
The entries of CatPred-DB datasets have complete enzymatic annotations, including kinetic parameters, organism identity, enzyme sequence, and substrate SMILES string, with taxonomy and sequence information verified via NCBI Taxonomy database \cite{schoch2020ncbi} and UniProt database \cite{uniprot2025uniprot}.
For each enzyme–substrate pair, a single representative kinetic value was assigned by selecting the maximum $k_{cat}$ and the geometric mean of $K_m$ and $K_i$, ensuring consistency and stability during machine-learning model training.
Finally, there are 23,197, 41,174 and 11,929 measurements in $k_{cat}$ dataset, $K_m$ dataset, and $K_i$ dataset of CatPred-DB, respectively.
The authors of CatPred-DB literature \cite{boorla2025catpred} divided each dataset into training set, validation set and test set according to the hold-out method with a ratio of 80\%, 10\% and 10\% respectively.
Since the maximum percentage sequence identity (Max. \% seq. id. cutoff) of the held-out test set to training sequences is 100\%, it was used for in-distribution performance evaluation in this work.
The authors further divided each held-out test set into four test subsets with different sequence identity cutoff values (99\%, 80\%, 60\%, and 40\%), which were used for out-of-distribution performance evaluation in this work.

\subsubsection{Enzyme-substrate pairing dataset}\label{method_pair_data}

The ESP database from a previous work \cite{kroll2023esp} was adopted as the benchmark dataset for enzyme-substrate pairing prediction tasks.
This dataset consists of enzyme-substrate pairs curated from GO annotation database \cite{dimmer2012uniprot}.
18,351 enzyme-substrate pairs with experimental evidence for binding and 274,030 pairs with phylogenetically inferred evidence for binding were extracted, and a molecular fingerprint-based similarity score was used to randomly sample non-binding enzyme-small molecule pairs, which resulted in the final ESP database of 784,396 entries, including 212,111 positive samples (binding pairs) and 572,282 negative samples (non-binding pairs).
The authors of ESP database literature \cite{kroll2023esp} split this dataset into training/validation/test set with 765,639/5,421/13,336 samples.
Since the enzyme sequences in the test set have a maximum percentage sequence identity of 80\% to training sequences, it was used for out-of-distribution performance evaluation in this work (referred to as "Overall" test set).
The authors further divided the test set into three test subsets with different intervals of maximum percentage sequence identity (0-40\%, 40-60\%, and 60-80\%), which were used for additional out-of-distribution evaluation in this work.

\subsubsection{Enzymatic active site dataset}\label{method_site_data}

The SwissProt E-RXN ASA dataset from a previous work \cite{wang2024multi} was adopted as the benchmark dataset for enzymatic active site annotation tasks.
This dataset consists of enzyme-reaction pairs curated from Swiss-Prot \cite{boutet2007uniprotkb} and ECReact dataset \cite{probst2022biocatalysed}.
The authors of SwissProt E-RXN ASA dataset literature \cite{wang2024multi} constructed this dataset by selecting SwissProt entries with EC numbers and annotated functional sites, retaining enzymes $\leq600$ amino acids with AlphaFold2-predicted structures and standardized active site labels.
The entries with experimentally verified structures were adopted as the validation set and test set, and the training sequences sharing more than 80\% sequence identity with those in validation/test sets via CD-HIT \cite{fu2012cd} were removed.
Subsequently, enzymatic reactions from ECReact (62,222 entries) were matched to the above divided SwissProt enzyme active site data via EC numbers.
Finally, there are 102,944/4,711/892 enzyme–reaction pairs for training/validation/test sets, where each pair includes the enzyme sequence, predicted structure, reaction SMILES, and active site annotations.
Since the enzyme sequences in the test set have a maximum percentage sequence identity of 80\% to training sequences, it was used for out-of-distribution performance evaluation in this work (referred to as "Overall" test set).
The authors further divided the test set into five test subsets with different intervals of maximum percentage sequence identity (0-40\%, 40-50\%, 50-60\%, 60-70\%, and 70-80\%), which were used for more comprehensive out-of-distribution evaluation in this work.

The MCSA E-RXN CSA dataset is another enzymatic active site dataset of the previous work \cite{wang2024multi}, where the entries in it have high quality annotations curated from the Mechanism and Catalytic Site Atlas dataset \cite{ribeiro2018mechanism}.
The raw test set of MCSA E-RXN CSA dataset was adopted in this work to demonstrate the interpretability of OmniESI's attention weights through zero-shot inference, where the enzymes in it had maximum 80\% sequence identity to those in SwissProt E-RXN ASA training set.

\subsubsection{Enzyme mutational effect dataset}\label{method_mutation_data}

To evaluate the predictive ability of OmniESI for enzyme mutational effects, we created a single-point mutational effect dataset and two double-point mutational effect datasets from the deep mutational scanning experiments of CTX-M $\beta$-lactamase \cite{judge2023mapping}.

The single-point mutational effect dataset was derived from deep mutational scanning results at 17 active sites in CTX-M $\rm \beta$-lactamase, containing fitness measurements against three substrates: cefotaxime, ceftazidimi and ampicillin.
Mutations with the relative fitness to the wild-type residue greater than 0 were considered as positive samples (beneficial mutations), while those with the relative fitness to the wild-type residue less than 0 were considered as negative samples (harmful mutations), thereby resulting in 93 positive samples and 866 negative samples.
To fairly evaluate the prediction performance under highly unbalanced sample distributions, we constructed the test set by selecting 20\% of the positive samples and an equal number of negative samples.
The validation set was formed similarly, with the rest samples used as training set.

The double-point mutational effect datasets were curated from deep mutational scanning results at 17 active sites in CTX-M $\rm \beta$-lactamase against two substrates, namely cefotaxime and ampicillin.
Based on the annotations of positive and negative epistasis, the double-point mutational effect dataset for cefotaxime and ampicillin was created respectively.
There are 38,880 positive and 599 negative epistasis samples in the double-point mutational effect dataset for cefotaxime, while 32,438 positive and 1,971 negative epistasis samples in the double-point mutational effect dataset for ampicillin.
To fairly evaluate the prediction performance under highly unbalanced sample distributions, we constructed the test set for both datasets by selecting 20\% of the negative epistasis samples along with an equal number of positive epistasis samples.
The validation sets for both datasets were created using the same approach, while the remaining samples were assigned to the training sets.

\subsection{Architecture overview}

OmniESI is composed of four modules, namely enzyme-substrate encoding, bidirectional conditional feature modulation (BCFM), catalysis-aware conditional feature modulation (CCFM), and task-specific decoding.
Given the input enzyme amino-acid sequence $\textbf{I}_{\mathrm{seq}}$ with $\rm N$ residues and substrate 2D graph $\textbf{I}_{\mathrm{graph}}$ with $\rm M$ atoms, ESM-2 (650M) and graph convolutional network (GCN) are employed to encode them into protein and molecular domain features, respectively. 
The encoded features are then processed through two multi-layer perceptrons (MLPs) to ensure consistent output dimensions:
\begin{equation}
\begin{aligned}
\rm \textbf{F}_{p} &= \rm MLP(ESM(\textbf{I}_{\mathrm{seq}})) \\
\rm \textbf{F}_{m} &= \rm MLP(GCN(\textbf{I}_{\mathrm{graph}}))
\end{aligned}
\label{eq:encode_esm_gcn}
\end{equation}
where $\rm \textbf{F}_{p} \in \mathbb{R}^{N\times d}$ and $\rm \textbf{F}_{m} \in \mathbb{R}^{M\times d}$ denote the enzyme and substrate features in protein-molecule domain, respectively, $\rm d$ stands for the hidden dimension.

The conditional network I (i.e. BCFM module) $\mathcal{F}(\cdot)$ is then applied, where paired $\rm \textbf{F}_p$ and $\rm \textbf{F}_m$ serve as conditions for each other to further map the features into the enzyme-substrate domain:
\begin{equation}
\begin{aligned}
\rm \textbf{F}_{e} &= \rm \mathcal{F}^{p \rightarrow e}(\textbf{F}_{p} | \textbf{F}_{m}) \\
\rm \textbf{F}_{s} &= \rm \mathcal{F}^{m \rightarrow s}(\textbf{F}_{m} | \textbf{F}_{p})
\end{aligned}
\label{eq:BCFM}
\end{equation}
where $^{* \rightarrow *}$ represents the direction of domian transition using the conditional network.
$\rm \textbf{F}_{e} \in \mathbb{R}^{N\times d}$ and $\rm \textbf{F}_{s} \in \mathbb{R}^{M\times d}$ denote the features projected into enzyme and substrate domains, respectively.

Subsequently, the conditional network II $\mathcal{G}(\cdot)$ (i.e. CCFM module) derives the initial interaction representation $\rm \textbf{F}_{inter}^{init} \in \mathbb{R}^{1\times d}$ by integrating $\rm \textbf{F}_{e}$ and $\rm \textbf{F}_{s}$, employing it as a condition for $\rm \textbf{F}_{e}$ and $\rm \textbf{F}_{s}$ to capture features within the catalysis-aware domain:
\begin{equation}
\begin{aligned}
\rm \textbf{F}_{inter}^{init} &= \rm \mathcal{A}(\textbf{F}_{e}, \textbf{F}_{s}) \\
\rm \textbf{F}_{e}^{cata} &= \rm \mathcal{G}^{e \rightarrow cata}(\textbf{F}_{e} | \textbf{F}_{inter}^{init})  \\
\rm \textbf{F}_{s}^{cata} &= \rm \mathcal{G}^{s \rightarrow cata}(\textbf{F}_{s} | \textbf{F}_{inter}^{init})
\end{aligned}
\label{eq:CCFM}
\end{equation}
where $\rm \textbf{F}_{e}^{cata} \in \mathbb{R}^{N\times d}$ and $\rm \textbf{F}_{s}^{cata} \in \mathbb{R}^{M\times d}$ denote the catalysis-aware enzyme and substrate representations, respectively. $\rm \mathcal{A}$ refers to the aggregation network.

Ultimately, the catalysis-aware interaction representation $\rm \textbf{F}_{inter}^{cata} \in \mathbb{R}^{1\times d}$ is obtained from $\rm \textbf{F}_{\text{e}}^{cata}$ and $\rm \textbf{F}_{\text{s}}^{cata}$ and decoded by the MLP to reveal the task-specific targets it contains:
\begin{equation}
\begin{aligned}
\rm \textbf{F}_{inter}^{cata} &= \rm \mathcal{P}_{\text{avg}}(\textbf{F}_{\text{e}}^{cata}) \oplus \mathcal{P}_{\text{avg}}(\textbf{F}_{\text{s}}^{cata}) \\
\rm \textbf{Y}_* &= \rm \text{MLP}_*(\textbf{F}_{inter}^{cata}), \quad * = \{\text{cls, reg}\}
\end{aligned}
\label{eq:decode_2}
\end{equation}
where $\mathcal{P}_{\text{avg}}$ represents average pooling, $\oplus$ denotes element-wise addition, $*$ indicates the task type i.e., classification or regression, and $\rm \textbf{Y}_{*}$ is the output result.

\subsection{Bidirectional conditional feature modulation}

The proposed BCFM module includes two core blocks: the poly focal perception (PFP) block, which extracts multi-receptive field features, and the two-sided conditioning (TC) block, which further guides the model to emphasize enzyme- and substrate-specific features. 
The representation learning process of enzymes and their substrates follows a bidirectional symmetric manner: initially, the substrate is used as a condition to support enzyme feature extraction, and then the enzyme acts as a condition to facilitate substrate feature extraction.

\subsubsection{Poly focal perception block}
The PFP block is designed on top of the poly focal attention (PFA) operation, which adopts the recently popular large kernel convolution paradigm \cite{yu2024inceptionnext, cai2024pki}.
By cascading multiple large kernel depthwise separable convolutions (DSCNN) \cite{chollet2017depthwise}, it enables the extraction of fine-grained features across diverse receptive fields, ensuring comprehensive contextual representation for both enzymes and substrates. 
In data-scarce scenarios, this approach exhibits faster convergence and a reduced risk of overfitting compared to self-attention according to previous studies \cite{li2023lsknet, hou2024conv2former}.

Specifically, let $\rm \textbf{X} \in \mathbb{R}^{n\times d}$ represent the input feature sequence, which could be either $\rm \textbf{F}_p$ or $\rm \textbf{F}_m$, where $\rm n$ denotes the sequence length.
We first employ a linear layer $\rm \phi(\cdot)$ to transform $\textbf{X}$ into the initial features $\rm \textbf{X}_0 \in \mathbb{R}^{n\times d}$.
Subsequently, multi-scale features are extracted through $\rm L$ cascaded DSCNN.
For $l = 1, 2, \dots, \text{L}$, the feature at the $l$-th layer $\textbf{X}_{l} \in \mathbb{R}^{\rm n\times d}$ is formulated as:
\begin{equation}
\textbf{X}_{l} = \text{ReLU}(\text{BN}(\text{DSCNN}_{\text{k}_{l}}(\textbf{X}_{l-1})))
\label{eq:pfp_1}
\end{equation}
where BN denotes the 1D batch normalization operator, ReLU stands for the activation function, and $\text{k}_{l}$ refers to the kernel size at the $l$-th DSCNN. 
Following that, the forward process of PFA operation is functionalized as $\rm PFA(\cdot)$ by aggregating all $\rm L$ layers:
\begin{equation}
\rm PFA(\textbf{X}) = \textbf{X}_{1} \oplus \textbf{X}_{2} \oplus \textbf{X}_{3} \oplus \dots \oplus \textbf{X}_{\text{L}}
\label{eq:pfp_2}
\end{equation}

Finally, we construct the forward process of PFP block through PFA operation and other essential network elements:
\begin{equation}
\begin{aligned}
\rm \textbf{X}_{PFA} &=\rm PFA(LN_{1}(\textbf{X})) + \textbf{X} \\
\rm \textbf{X}_{PFP} &=\rm FFN(LN_{2}(\textbf{X}_{PFA})) + \textbf{X}_{PFA}
\end{aligned}
\label{eq:pfp_3}
\end{equation}
where $\rm \textbf{X}_{\rm PFA} \in \mathbb{R}^{\text{n}\times d}$ and $\rm \textbf{X}_{\rm PFP} \in \mathbb{R}^{\text{n}\times d}$ denotes the output of the PFA operation and PFP block, respectively, FFN represents the feed-forward network, while $\rm LN_1$ and $\rm LN_2$ refer to layer normalization operations with non-shared parameters.

\subsubsection{Two-sided conditioning block}

The specificity of enzymatic reactions \cite{ryan1989specificity} naturally forms enzyme-substrate pairs as binary tuple data.
Similar to the paired images and captions in multimodal learning \cite{radford2021clip, ramesh2021dalle, ruiz2023dreambooth}, these enzyme-substrate pairs can serve as conditions for each other during the representation learning process.
To better leverage the conditional relationship, we propose the TC block, which generates enzyme-side or substrate-side conditional embeddings, to guide the PFP block in extracting task-relevant features.
Since enzyme-side and substrate-side conditioning are symmetrical, we take enzyme-side conditioning process as an example to demonstrate how the substrate is employed to generate conditional embeddings that direct the feature learning process of enzyme.

Specifically, we first obtain the global representation of the substrate $\rm \textbf{F}_{m}^{avg} \in \mathbb{R}^{1\times d}$ through average pooling and treat it as a query vector to derive the importance distribution of enzyme residues $\rm \textbf{A}_{p} \in \mathbb{R}^{N\times 1}$:
\begin{equation}
\rm \mathbf{A}_{p} = \text{Softmax}\left( \frac{\phi_{q}(\textbf{F}_{m}^{\text{avg}})  \phi_{k}(\mathbf{F}_{\text{p}})^{\top}}{\sqrt{d}} \right)
\label{eq:scm_1}
\end{equation}
where $\rm \phi_{q}(\cdot)$ and $\rm \phi_{k}(\cdot)$ represent the linear transformations of the query and key, respectively.
Accordingly, the substrate-guided enzyme conditional embedding $\rm \mathbf{C}_{p} \in \mathbb{R}^{1\times d}$ is obtained based on $\rm \textbf{A}_{p}$:
\begin{equation}
\rm \mathbf{C}_{p} = \mathcal{P}_{\text{avg}}(\mathbf{F}_{p} \odot \mathbf{A}_{p}) = \frac{1}{N} \sum_{i=1}^{N} ( f_i \cdot a_i)
\label{eq:scm_2}
\end{equation}
where $\rm f_i \in \mathbf{F}_{p}$ and $\rm a_i \in \mathbf{A}_{p}$ denote the i-th residue feature and corresponding attention score in enzyme, $\odot$ indicate the element-wise multiplication. 
Subsequently, we utilize $\rm \mathbf{C}_p$ to guide the enzyme feature extraction network in actively capturing enzyme-specific features, thereby achieving the transformation from the protein domain to the enzyme domain.
Specifically, we employ adaptive layer normalization, which follows a soft yet effective condition-incorporating paradigm.
Based on this, we directly regress the parameters $\rm \gamma \in \mathbb{R}^{d}$ and $\rm \beta \in \mathbb{R}^{d}$ required for layer normalization \cite{perez2018adaLN} in PFP block from $\rm \mathbf{C}_p$. 
Therefore, PFP block with the enzyme-side conditioning, denoted as $\rm PFP^{cond}(\cdot)$, is defined as:
\begin{equation}
\begin{aligned}
\rm PFP^{cond}(\mathbf{F}_{p}|\mathbf{F}_{m})&=\rm PFP(\mathbf{F}_{p}|LN_{1}=LN_{1}^{cond}, LN_{2}=LN_{2}^{cond})\\
\rm LN_{1}^{cond}&=\rm LN_{1}(\gamma=\gamma_{1}, \beta=\beta_{1})\\
\rm LN_{2}^{cond}&=\rm LN_{2}(\gamma=\gamma_{2}, \beta=\beta_{2}) \\
\rm \gamma_{1}, \beta_{1}, \gamma_{2}, \beta_{2} &= \rm \mathcal{S}(MLP(\mathbf{C}_p))
\end{aligned}
\label{eq:scm_3}
\end{equation}
where $\rm LN^{cond}$ represents the layer normalization layer whose $\gamma$ and $\beta$ parameters are generated by the conditional embedding, $\mathcal{S}(\cdot)$ indicates the operation of splitting along the hidden dimension.

Similarly, the output of the substrate feature can be symmetrically obtained via the PFP block with substrate-side conditioning, denoted as $\rm PFP^{cond}(\mathbf{F}_{m}|\mathbf{F}_{p})$.

\subsubsection{Bidirectional feature modulation}
Utilizing the above PFP and TC block, a bidirectional feature modulation paradigm is established:
\begin{equation}
\begin{aligned}
\rm \textbf{F}_{e} &= \rm \mathcal{F}^{p \rightarrow e}(\textbf{F}_{p} | \textbf{F}_{m}) =  PFP^{cond}(\mathbf{F}_{p}|\mathbf{F}_{m})\\
\rm \textbf{F}_{s} &= \rm \mathcal{F}^{m \rightarrow s}(\textbf{F}_{m} | \textbf{F}_{p}) = PFP^{cond}(\mathbf{F}_{m}|\mathbf{F}_{p})
\end{aligned}
\label{eq:bfm}
\end{equation}
The modulated features, $\mathbf{F}_e$ and $\mathbf{F}_s$, respectively highlight enzyme-specific and substrate-specific features.
Notably, after completing the forward process of the PFP block on each side, multi-head cross-attention \cite{vaswani2017attention} is adopted to effectively use the modulated features as the key and value, updating the features serving as the condition.

\subsection{Catalysis-aware conditional feature modulation}

The initial aggregated representation formed by concatenating the representations of an enzyme and its substrate modulated by BCFM module represents rough enzyme-substrate interaction patterns.
Therefore, we devise CCFM module that leverages this initial representation as a conditional embedding, enabling fine-grained re-weighting of reaction-related residues in the enzyme and atoms in the substrate, thereby obtaining interaction-aware features.

\subsubsection{Interaction conditioning}

The main process of interaction conditioning involves obtaining the global representations of the enzyme and substrate, then aggregating them to derive the conditional embedding that represents the initial reaction representation:
\begin{equation}
\begin{aligned}
\rm \textbf{F}_{inter}^{init} = \phi(\rm Concat(\mathcal{P}_{avg}(MLP(\textbf{F}_e)), \mathcal{P}_{avg}(MLP(\textbf{F}_s))))\\
\end{aligned}
\label{eq:ccfm_1}
\end{equation}
where $\rm \textbf{F}_{inter}^{init} \in \mathbb{R}^{1\times d}$ represents the conditional embedding, $\phi(\cdot)$ denotes the linear layer ensuring dimensional consistency, $\rm Concat(\cdot)$ refers to the concatenation.

\subsubsection{Catalysis-aware feature modulation}

We adopt the conditional embedding $\rm \textbf{F}_{inter}^{init}$ as the query to exploit which residue or atom is crucial under the perspective of interaction pattern, which is formulated as:
\begin{equation}
\begin{aligned}
\rm \mathbf{A}_{e}^{cata} = \text{Softmax}\left( \frac{\phi_{q}(\textbf{F}_{inter}^{init})  \phi_{k}(\mathbf{F}_{\text{e}})^{\top}}{\sqrt{d}} \right) \\
\rm \mathbf{A}_{s}^{cata} = \text{Softmax}\left( \frac{\phi_{q}(\textbf{F}_{inter}^{init})  \phi_{k}(\mathbf{F}_{\text{s}})^{\top}}{\sqrt{d}} \right) 
\label{eq:ccfm_2}
\end{aligned}
\end{equation}
where $\rm \mathbf{A}_{e}^{cata} \in \mathbb{R}^{N\times 1}$ and $\rm \mathbf{A}_{s}^{cata} \in \mathbb{R}^{M\times 1}$ denote the catalysis-aware attention scores for enzyme and substrate, respectively.
These attention scores are then applied to highlight residues and atoms directly involved in the interaction, thus achieving the features transformations from enzyme-substrate domain to catalysis-aware domain:
\begin{equation}
\begin{aligned}
\rm \textbf{F}_{e}^{cata} &= \rm \mathcal{G}^{e \rightarrow inter}(\textbf{F}_{e} | \textbf{F}_{inter}^{init}) =  \mathbf{F}_{e} \odot \mathbf{A}_{e}^{cata}  \\
\rm \textbf{F}_{s}^{cata} &= \rm \mathcal{G}^{s \rightarrow inter}(\textbf{F}_{s} | \textbf{F}_{inter}^{init}) =  \mathbf{F}_{s} \odot \mathbf{A}_{s}^{cata}
\end{aligned}
\label{eq:CCFM}
\end{equation}
The catalysis-aware features $\rm \textbf{F}_{e}^{cata}$ and $\rm \textbf{F}_{s}^{cata}$ highlight the enzyme residues and substrate atoms involved in catalysis process, thereby resulting in a fine-grained enzyme-substrate interaction pattern.
For the implementation, the calculations of $\rm \mathbf{A}_{e}^{cata}$ and $\rm \mathbf{A}_{s}^{cata}$ adopt a multi-head mechanism to capture more expressive multi-view features, with the number of heads empirically set to 4.

\subsection{Evaluation metrics}

\subsubsection{Metrics for regression tasks}

To evaluate the regression performance of enzyme kinetic parameter prediction, the coefficient of determination ($\rm R^{2}$) and the Mean Absolute Error (MAE) are calculated as follows:

\begin{equation}
\rm R^2 = 1 - \frac{\sum_{i=1}^{n} (y_{\text{label}}^{i} - y_{\text{pred}}^{i})^2}{\sum_{i=1}^{n} (y_{\text{label}}^{i} - \bar{y}_{label})^2}
\end{equation}

\begin{equation}
\rm MAE =\frac{1}{n}\sum_{i=1}^{n}{\left\lvert y_{\text{label}}^{i} - y_{\text{pred}}^{i}  \right\rvert}
\end{equation}

where $\rm y_{\text{label}}^{i}$ and $\rm y_{\text{pred}}^{i}$ refer to the actual target and predicted values, respectively, while $\rm \bar{y}_{\text{label}}$ indicates the average of target values.
The number of samples is denoted as $\rm n$.

\subsubsection{Metrics for classification tasks}

To evaluate the classification performance of enzyme-substrate pairing prediction, enzyme mutational effect prediction, and enzymatic active site annotation, the Accuracy, F1 score, Matthews Correlation Coefficient (MCC), Area Under the Receiver Operating Characteristic Curve (AUROC), and Area Under the Precision-Recall Curve (AUPRC) are adopted.

In binary classification tasks, four possible prediction outcomes are true positives (TP), true negatives (TN), false positives (FP), and false negatives (FN). 
Afterwards, Accuracy and F1-score can be respectively formulated as follows:
\begin{equation}
\rm Accuracy = \frac{TP + TN}{TP + TN + FP + FN}
\label{eq:acc}
\end{equation}
\begin{equation}
\rm Precision = \frac{TP}{TP + FP}
\label{eq:precision}
\end{equation}
\begin{equation}
\rm Recall = \frac{TP}{TP + FN}
\label{eq:recall}
\end{equation}
\begin{equation}
\rm F1~score = 2 \times \frac{Precision \times Recall}{Precision + Recall}
\label{eq:recall}
\end{equation}

Moreover, MCC incorporates all elements of the confusion matrix, i.e., TP, TN, FP, and FN, offering a robust evaluation for class-imbalanced classification tasks:
\begin{equation}
\rm MCC = \frac{TP \times TN - FP \times FN}{\sqrt{(TP + FP)(TP + FN)(TN + FP)(TN + FN)}}
\label{eq:mcc}
\end{equation}

In addition, AUPRC and AUROC are used to further evaluate the binary classification performance across different confidence thresholds. 
AUPRC evaluates the trade-off between Precision and Recall by plotting the Precision-Recall curve and calculating the area under the curve, making it particularly suitable for imbalanced datasets with a small proportion of positive samples.
AUROC, in contrast, measures the area under the receiver operating characteristic curve, which plots the true positive rate against the false positive rate, offering a more comprehensive assessment of a classifier’s overall discriminative ability.

\subsection{Implementation}
The OmniESI model is implemented using Python 3.8 and PyTorch 1.12.1.
For substrate encoding, the GCN network provided by DGL \cite{wang2019deep} and DGL-LifeSci \cite{li2021dgl} was utilized.
For enzyme encoding, the ESM-2-650M pretrained weights \cite{lin2023esm} were employed.
Additionally, the implementation of conditional networks primarily leverages the timm \footnote{https://huggingface.co/timm} and transformers \footnote{https://huggingface.co/docs/transformers/index} libraries.
During training, the Adam optimizer was adopted, and Distributed Data Parallel (DDP) \footnote{https://docs.pytorch.org/docs/stable/notes/ddp.html} was used to accelerate the training process. 
The model training was performed in parallel using four NVIDIA A800 GPUs.
The performance of OmniESI is not sensitive to specific architecture parameter choices.
A unified set of architecture parameters was used across all downstream tasks, with only batch size and learning rate adjusted appropriately according to the scale of different datasets (Supplementary \ref{SI_Hyperparameters}). 
The inference stage of OmniESI can be efficiently executed on a single NVIDIA RTX 3090 GPU.
The required inputs are only the enzyme sequences and their substrates' SMILES strings.
PyMOL \cite{delano2002pymol} and RDKit \footnote{https://www.rdkit.org/} are utilized for visualizing protein structures and the attention distribution of substrates.

\section*{Data and code availability}

The enzyme kinetic parameter dataset, the enzyme-substrate pairing dataset, and the enzymatic active site dataset are freely available from previous studies \cite{boorla2025catpred,kroll2023esp,wang2024multi}.
Relevant data, code, and models of this work are available via GitHub at \url{https://github.com/Hong-yu-Zhang/OmniESI}.

\bibliography{sn-bibliography}

\newpage
\section*{Supplementary information}

\setcounter{section}{1}
\renewcommand{\thesubsection}{S\arabic{section}}
\subsection{Prediction performance for enzyme kinetic parameters}
\textbf{Bold} indicates the best performance. 
\underline{Underline} indicates the second-best.
The raw results of the other models are taken from a prior work \cite{boorla2025catpred}.

\begin{table}[h]
\renewcommand{\thetable}{S1}
\centering
\begin{tabular}{ccccccccccc}
\toprule
 & \multicolumn{2}{c}{ID - 100\%} & \multicolumn{2}{c}{OOD - 99\%} & \multicolumn{2}{c}{OOD - 80\%} & \multicolumn{2}{c}{OOD - 60\%} & \multicolumn{2}{c}{OOD - 40\%} \\
\cmidrule(lr){2-3} \cmidrule(lr){4-5} \cmidrule(lr){6-7} \cmidrule(lr){8-9} \cmidrule(lr){10-11}
Method & R2 $\uparrow$ & MAE $\downarrow$ & R2 $\uparrow$ & MAE $\downarrow$ & R2 $\uparrow$ & MAE $\downarrow$ & R2 $\uparrow$ & MAE $\downarrow$ & R2 $\uparrow$ & MAE $\downarrow$ \\
\midrule
DLKcat & 0.239 &1.465 &0.117 &1.523 &0.108 &1.518 &0.085 &1.583 &0.102 &1.556 \\
UniKP  & 0.591 &0.737 &0.372 &1.043 &0.349 &1.071 &0.302 &1.179 &0.260 &1.189 \\
$\rm CatPred$ & \underline{0.602} &\underline{0.714} &\underline{0.404} &\underline{0.987} &\underline{0.383} &\underline{1.029} &\underline{0.355} &\underline{1.124} &\textbf{0.365} &\textbf{1.131} \\
$\rm \textbf{OmniESI}$ &\textbf{0.638} &\textbf{0.670} &\textbf{0.409} &\textbf{0.956} &\textbf{0.401} &\textbf{0.989} &\textbf{0.356} &\textbf{1.062} &\underline{0.342} &\underline{1.142}  \\
\bottomrule
\end{tabular}
\caption{Performance comparison of different models on CatPredDB-$k_{cat}$ dataset.}
\label{tables:kcat}
\end{table}

\begin{table}[h]
\renewcommand{\thetable}{S2}
\centering
\begin{tabular}{ccccccccccc}
\toprule
 & \multicolumn{2}{c}{ID - 100\%} & \multicolumn{2}{c}{OOD - 99\%} & \multicolumn{2}{c}{OOD - 80\%} & \multicolumn{2}{c}{OOD - 60\%} & \multicolumn{2}{c}{OOD - 40\%} \\
\cmidrule(lr){2-3} \cmidrule(lr){4-5} \cmidrule(lr){6-7} \cmidrule(lr){8-9} \cmidrule(lr){10-11}
Method & R2 $\uparrow$ & MAE $\downarrow$ & R2 $\uparrow$ & MAE $\downarrow$ & R2 $\uparrow$ & MAE $\downarrow$ & R2 $\uparrow$ & MAE $\downarrow$ & R2 $\uparrow$ & MAE $\downarrow$ \\
\midrule
DLKcat &0.440 &1.229 &0.383 &1.261 &0.392 &1.259 &0.397 &1.280 &0.353 &1.291 \\
UniKP  &0.594 &0.606 &0.493 &0.688 &0.496 &0.708 &0.499 &0.731 &0.449 &0.800 \\
$\rm CatPred$  &\underline{0.642} &\underline{0.558} &\underline{0.533} &\underline{0.653} &\underline{0.540} &\underline{0.665} &\textbf{0.539} &\textbf{0.693} &\textbf{0.475} &\underline{0.771} \\
$\rm \textbf{OmniESI}$ &\textbf{0.654} &\textbf{0.541} &\textbf{0.541} &\textbf{0.639} &\textbf{0.549} &\textbf{0.655} &\underline{0.534} &\underline{0.694} &\underline{0.472} &\textbf{0.765} \\
\bottomrule
\end{tabular}
\caption{Performance comparison of different models on CatPredDB-$K_{m}$ dataset.}
\label{tables:km}
\end{table}

\begin{table}[h]
\renewcommand{\thetable}{S3}
\centering
\begin{tabular}{ccccccccccc}
\toprule
 & \multicolumn{2}{c}{ID - 100\%} & \multicolumn{2}{c}{OOD - 99\%} & \multicolumn{2}{c}{OOD - 80\%} & \multicolumn{2}{c}{OOD - 60\%} & \multicolumn{2}{c}{OOD - 40\%} \\
\cmidrule(lr){2-3} \cmidrule(lr){4-5} \cmidrule(lr){6-7} \cmidrule(lr){8-9} \cmidrule(lr){10-11}
Method & R2 $\uparrow$ & MAE $\downarrow$ & R2 $\uparrow$ & MAE $\downarrow$ & R2 $\uparrow$ & MAE $\downarrow$ & R2 $\uparrow$ & MAE $\downarrow$ & R2 $\uparrow$ & MAE $\downarrow$ \\
\midrule
DLKcat &0.508 &1.977 &0.348 &1.708 &0.237 &1.626 &0.260 &1.604 &0.298 &1.621\\
UniKP &0.600 &0.937 &0.443 &1.059 &0.359 &1.115 &0.345 &1.103 &0.400 &1.074 \\
$\rm CatPred$ &\underline{0.632} &\underline{0.876} &\underline{0.473} &\underline{0.988} &\underline{0.388} &\underline{1.036} &\underline{0.362} &\underline{1.036} &\underline{0.401} &\underline{1.029} \\
$\rm \textbf{OmniESI}$ &\textbf{0.663} &\textbf{0.834} &\textbf{0.543} &\textbf{0.932} &\textbf{0.465} &\textbf{0.982} &\textbf{0.463} &\textbf{0.987} &\textbf{0.445} &\textbf{1.019} \\
\bottomrule
\end{tabular}
\caption{Performance comparison of different models on CatPredDB-$K_{i}$ dataset.}
\label{tables:ki}
\end{table}

\newpage
\newpage
\setcounter{section}{2}
\renewcommand{\thesubsection}{S\arabic{section}}
\subsection{Prediction performance for enzyme-substrate pairs}
\textbf{Bold} indicates the best performance.
The raw results of the other models are taken from a prior work \cite{kroll2023esp}.

\begin{table}[!htbp]
\renewcommand{\thetable}{S4}
\centering
\begin{tabular}{cccc}
\toprule
{Model} & {MCC ↑} & {AUROC ↑} & {Accuracy ↑} \\
\midrule
ESP & 0.780 & 0.956 & 0.915 \\
\textbf{OmniESI} & \textbf{0.859} & \textbf{0.972} & \textbf{0.946} \\
\bottomrule
\end{tabular}
\caption{Performance comparison of different models on "Overall" test set of ESP database.}
\label{tables:pair_overall}
\end{table}

\begin{table}[h]
\renewcommand{\thetable}{S5}
\centering
\resizebox{\textwidth}{!}{
\begin{tabular}{ccccccccccc}
\toprule
 & \multicolumn{3}{c}{0$\sim$40\%} & \multicolumn{3}{c}{40$\sim$60\%} & \multicolumn{3}{c}{60$\sim$80\%} \\
\cmidrule(lr){2-4} \cmidrule(lr){5-7} \cmidrule(lr){8-10}
Method & MCC $\uparrow$ & AUROC $\uparrow$ & Accuracy $\uparrow$ & MCC $\uparrow$ & AUROC $\uparrow$ & Accuracy $\uparrow$ & MCC $\uparrow$ & AUROC $\uparrow$ & Accuracy $\uparrow$ \\
\midrule
ESP &0.720 &0.930 &0.890 &0.830 &0.970 &0.930 &0.880 &\textbf{0.990} &0.950 \\
\textbf{OmniESI} &\textbf{0.761} &\textbf{0.948} &\textbf{0.910} &\textbf{0.899} &\textbf{0.977} &\textbf{0.961} &\textbf{0.914} &0.987 &\textbf{0.967} \\
\bottomrule
\end{tabular}}
\caption{Performance comparison of different models on three test subsets with different sequence identity intervals of ESP database.}
\label{tables:pair_subset}
\end{table}

\newpage
\setcounter{section}{3}
\renewcommand{\thesubsection}{S\arabic{section}}
\subsection{Prediction performance for enzymatic active sites}

\textbf{Bold} indicates the best performance.
\underline{Underline} indicates the second-best.
The raw results of the other models are taken from a prior work \cite{wang2024multi}, where the AEGAN model didn't remove 225 test samples overlapping with its training set and the BLASTp adopted the entire SwissProt as sequence alignment database.

\begin{table}[h]
\renewcommand{\thetable}{S6}
\centering
\begin{tabular}{ccc}
\toprule
{Method} & {F1 $\uparrow$} & {MCC $\uparrow$} \\
\midrule
SchrodingerSiteMap & 0.122 & 0.110 \\
AEGAN & 0.222 & 0.245 \\
BLASTp & 0.704 & 0.709 \\
EasIFA-ESM & \underline{0.792} & \underline{0.801} \\
EasIFA-SaProt & 0.787 & 0.797 \\
\textbf{OmniESI} & \textbf{0.827} & \textbf{0.823} \\
\bottomrule
\end{tabular}
\caption{Performance comparison of different models on "Overall" test set of SwissProt E-RXN ASA dataset.}
\label{tables:site_overall}
\end{table}

\begin{table}[h]
\renewcommand{\thetable}{S7}
\centering
\begin{tabular}{cccccc}
\toprule
Method & 0$\sim$40\% & 40$\sim$50\% & 50$\sim$60\% & 60$\sim$70\% & 70$\sim$80\% \\
\midrule
SchrodingerSiteMap &0.114 &0.125 &0.135 &0.130 &0.123 \\
AEGAN &0.229 &0.206 &0.219 &0.227 &0.218 \\
BLASTp &0.509 &0.743 &0.840 &0.882 &\underline{0.923} \\
EasIFA-ESM &0.654 &\textbf{0.836} &0.892 &\underline{0.923} &0.915 \\
EasIFA-Saprot &\underline{0.661} &0.804 &\underline{0.902} &0.892 &0.916 \\
\textbf{OmniESI} &\textbf{0.695} &\underline{0.826} &\textbf{0.907} &\textbf{0.933} &\textbf{0.954 }\\
\bottomrule
\end{tabular}
\caption{F1-score comparison of different models on five test subsets with different sequence identity intervals of SwissProt E-RXN ASA dataset.
}
\label{tables:site_f1}
\end{table}

\begin{table}[h]
\renewcommand{\thetable}{S8}
\centering
\begin{tabular}{cccccc}
\toprule
Method & 0$\sim$40\% & 40$\sim$50\% & 50$\sim$60\% & 60$\sim$70\% & 70$\sim$80\% \\
\midrule
SchrodingerSiteMap &0.101 &0.110 &0.122 &0.118 &0.117 \\
AEGAN &0.252 &0.225 &0.242 &0.248 &0.249 \\
BLASTp &0.516 &0.747 &0.840 &0.885 &\underline{0.927} \\
EasIFA-ESM &0.671 &\textbf{0.843} &\underline{0.894} &\underline{0.928} &0.919 \\
EasIFA-Saprot &\underline{0.679} &0.810 &\textbf{0.904} &0.898 &0.922 \\
\textbf{OmniESI} &\textbf{0.692} &\underline{0.820} &\textbf{0.904} &\textbf{0.931} &\textbf{0.953} \\
\bottomrule
\end{tabular}
\caption{MCC comparison of different models on five test subsets with different sequence identity intervals of SwissProt E-RXN ASA dataset.}
\label{tables:site_mcc}
\end{table}

\newpage
\setcounter{section}{4}
\renewcommand{\thesubsection}{S\arabic{section}}
\subsection{Prediction performance for enzyme mutational effects}
\label{fitness effects}

\textbf{Bold} indicates the best performance.

\begin{table}[!htbp]
\renewcommand{\thetable}{S9}
\centering
\begin{tabular}{cccccc}
\toprule
{Method} & {MCC $\uparrow$} & {AUPRC $\uparrow$} & {AUROC $\uparrow$} & {Accuracy $\uparrow$} & {F1 $\uparrow$}\\
\midrule
UniKP & 0.000 & 0.594 & 0.679 & 0.500 &0.500 \\
\textbf{OmniESI} & \textbf{0.752} & \textbf{0.928} & \textbf{0.898} & \textbf{0.861} &\textbf{0.839} \\
\bottomrule
\end{tabular}
\caption{Performance comparison of different models on the single-point mutational effect dataset.}
\label{tables:mutation_single}
\end{table}

\begin{table}[h]
\renewcommand{\thetable}{S10}
\centering
\begin{tabular}{cccccc}
\toprule
{Method} & {MCC $\uparrow$} & {AUPRC $\uparrow$} & {AUROC $\uparrow$} & {Accuracy $\uparrow$} & {F1 $\uparrow$}\\
\midrule
UniKP & 0.558 & 0.983 & 0.982 & 0.739 &0.792  \\
\textbf{OmniESI} & \textbf{0.692} & \textbf{0.996} & \textbf{0.995} & \textbf{0.824} &\textbf{0.850}\\
\bottomrule
\end{tabular}
\caption{Performance comparison of different models on the double-point mutational effect dataset for ampicillin.}
\label{tables:mutation_amp}
\end{table}

\begin{table}[h]
\renewcommand{\thetable}{S11}
\centering
\begin{tabular}{cccccc}
\toprule
{Method} & {MCC $\uparrow$} & {AUPRC $\uparrow$} & {AUROC $\uparrow$} & {Accuracy $\uparrow$} & {F1 $\uparrow$}\\
\midrule
UniKP & 0.730 & 0.986 & 0.988 & 0.847 & 0.868\\
\textbf{OmniESI} & \textbf{0.888} & \textbf{0.999} & \textbf{0.999} & \textbf{0.941} &\textbf{0.944}\\
\bottomrule
\end{tabular}
\caption{Performance comparison of different models on the double-point mutational effect dataset for cefotaxime.}
\label{tables:mutation_cefo}
\end{table}

\newpage
\setcounter{section}{5}
\renewcommand{\thesubsection}{S\arabic{section}}
\subsection{Module ablation experiments}

\textbf{Bold} indicates the best performance. 

\begin{table}[h]
\renewcommand{\thetable}{S12}
\centering
\begin{tabular}{ccccccccccc}
\toprule
 & \multicolumn{2}{c}{ID - 100\%} & \multicolumn{2}{c}{OOD - 99\%} & \multicolumn{2}{c}{OOD - 80\%} & \multicolumn{2}{c}{OOD - 60\%} & \multicolumn{2}{c}{OOD - 40\%} \\
\cmidrule(lr){2-3} \cmidrule(lr){4-5} \cmidrule(lr){6-7} \cmidrule(lr){8-9} \cmidrule(lr){10-11}
Method & R2 $\uparrow$ & MAE $\downarrow$ & R2 $\uparrow$ & MAE $\downarrow$ & R2 $\uparrow$ & MAE $\downarrow$ & R2 $\uparrow$ & MAE $\downarrow$ & R2 $\uparrow$ & MAE $\downarrow$ \\
\midrule
Baseline &0.572 &0.764 &0.344 &1.023 &0.327 &1.057 &0.301 &1.145 &0.237 &1.211 \\
w/ BCFM &0.624 &0.693 &0.375 &0.996 &0.357 &1.035 &0.319 &1.132 &0.287 &1.190 \\
w/ CCFM &0.613 &0.698 &0.352 &1.024 &0.340 &1.059 &0.313 &1.156 &0.246 &1.219 \\
$\rm \textbf{OmniESI}$ &\textbf{0.638} &\textbf{0.670} &\textbf{0.409} &\textbf{0.956} &\textbf{0.401} &\textbf{0.989} &\textbf{0.356} &\textbf{1.062} &\textbf{0.342} &\textbf{1.142}  \\
\bottomrule
\end{tabular}
\caption{Ablation experiments on CatPredDB-$k_{cat}$ dataset.}
\label{tables:ablation}
\end{table}

\setcounter{section}{6}
\renewcommand{\thesubsection}{S\arabic{section}}
\subsection{Parameter analysis}

\begin{table}[!htbp]
\renewcommand{\thetable}{S13}
\centering
\begin{tabular}{ccccccc}
\toprule
{Component} & {ESM-2}\footnotemark[1] & {GCN} & {BCFM} & {CCFM} & {MLPs} & {Total} \\
\midrule
Parameter Count &650.00M  & 0.11M & 0.90M & 0.10M & 0.95M &652.06M \\
Percentage &99.68\% &0.02\% &0.14\% & 0.02\% &0.15\% &- \\
\bottomrule
\end{tabular}
\footnotetext[1]{Parameters that are frozen during training}
\caption{Parameter analysis for each component of OmniESI.}
\label{tables:para_analysis}
\end{table}

\setcounter{section}{7}
\renewcommand{\thesubsection}{S\arabic{section}}
\subsection{Hyperparameters}\label{SI_Hyperparameters}

After performing step-by-step hyperparameter tuning, we found that the prediction performance of OmniESI is insensitive to the parameters of the architecture itself, and the unified architecture hyperparameters across all prediction tasks is shown in Table \ref{tables:model_hyper_parameters}.
Therefore, for different scales of datasets, we appropriately adjusted the training and optimization hyperparameters, namely batch size and learning rate.
First, there is one small-scale dataset, namely single-point mutational effect dataset, and its corresponding training configuration is shown in Table \ref{tables:small_data_hyper_parameters}.
Second, there are six medium-scale datasets, namely CatPred-DB $k_{cat}$ dataset, CatPred-DB $K_m$ dataset, CatPred-DB $K_i$ dataset, SwissProt E-RXN ASA dataset, and two double-point mutational effect datasets (cefotaxime and ampicillin), and their corresponding training configuration is shown in Table \ref{tables:medium_data_hyper_parameters}.
Third, there is one large-scale dataset, namely ESP dataset, and its corresponding training configuration is shown in Table \ref{tables:large_data_hyper_parameters}.

\begin{table}[!htbp]
\renewcommand{\thetable}{S14}
\centering
\begin{tabular}{cc}
\toprule
Hyperparameter & Value \\
\midrule
\textbf{Encoding} & \\
ESM embedding dim    & 1280  \\
GCN layers & 3   \\
GCN hidden dim & 128 \\
Enzyme output dim & 128 \\
Substrate output dim & 128 \\
\midrule
\textbf{BCFM} & \\
Hidden dim & 128 \\
Convolution layers   & 4  \\
Kernel size (Enzyme-side) & 5,9,13,17   \\
Kernel size (Substrate-side) & 3,5,7,9   \\
Cross-attention heads & 4 \\
Dropout rate & 0.1 \\
\midrule
\textbf{CCFM} & \\
Hidden dim & 128 \\
Attention heads & 4 \\
\bottomrule
\end{tabular}
\caption{Architecture hyperparameters across all prediction tasks.}
\label{tables:model_hyper_parameters}
\end{table}

\begin{table}[!htbp]
\renewcommand{\thetable}{S15}
\begin{tabular}{cc}
\toprule
Hyperparameter   & Value  \\
\midrule
Learning rate    & 1e-3 \\
Number of epochs & 50    \\
Batch size & 8 \\
Optimizer & Adam \\
Number of GPUs & 4 \\ 
\bottomrule
\end{tabular}
\caption{Training and optimization hyperparameters for small-scale datasets.}
\label{tables:small_data_hyper_parameters}
\end{table}

\begin{table}[!htbp]
\renewcommand{\thetable}{S16}
\begin{tabular}{cc}
\toprule
Hyperparameter   & Value  \\
\midrule
Learning rate    & 1e-4 \\
Number of epochs & 50    \\
Batch size & 32 \\
Optimizer & Adam \\
Number of GPUs & 4 \\ 
\bottomrule
\end{tabular}
\caption{Training and optimization hyperparameters for medium-scale datasets.}
\label{tables:medium_data_hyper_parameters}
\end{table}

\begin{table}[!htbp]
\renewcommand{\thetable}{S17}
\begin{tabular}{cc}
\toprule
Hyperparameter   & Value  \\
\midrule
Learning rate    & 1e-4 \\
Number of epochs & 50    \\
Batch size & 128 \\
Optimizer & Adam \\
Number of GPUs & 4 \\ 
\bottomrule
\end{tabular}
\caption{Training and optimization hyperparameters for large-scale datasets.}
\label{tables:large_data_hyper_parameters}
\end{table}

\end{document}